\documentclass[12pt]{article}
\usepackage{amssymb,amsmath,amsthm,amsfonts,amscd}
 \usepackage{graphicx}
\newcommand\be{\begin{equation}}
\newcommand\ee{\end{equation}}
\newcommand\bea{\begin{eqnarray}}
\newcommand\eea{\end{eqnarray}}
\newcommand\ket[1]{|#1\rangle}
\newcommand\bra[1]{\langle #1|}

\newcommand{\fatalpha}{{\bf \alpha \kern -0.44em \alpha}}
\newcommand{\fatsigma}{{\bf \sigma \kern -0.54em \sigma}}
\newcommand{\tpchi}{{\bf \chi \kern -0.35em \chi}}
\newcommand{\llambda}{{\bf \lambda \kern -0.45em \lambda}}

\renewcommand{\theequation}{\arabic{equation}}
\renewcommand{\theequation}{\thesection.\arabic{equation}}
\bibliography{plain}
\title{\bf Manipulating Multi-qudit Entanglement Witnesses by Using Linear Programming}\vspace{20mm}
\author{ M. A. Jafarizadeh$^{a,b,c}$
 \thanks{E-mail:jafarizadeh@tabrizu.ac.ir}  ,
 G. Najarbashi  $^{a,b}$
 \thanks{E-mail:najarbashi@tabrizu.ac.ir} , H. Habibian $^{a}$
 \thanks{E-mail:hesam-habibian@sicatechdec.com}
\\ $^a${\small Department of Theoretical Physics and Astrophysics,
Tabriz University, Tabriz 51664, Iran.} \\ $^b${\small Institute
for Studies in Theoretical Physics and Mathematics, Tehran
19395-1795, Iran.} \\ $^c${\small Research Institute for
Fundamental Sciences, Tabriz 51664, Iran.}} \pagebreak

 \newtheorem{thm}{Theorem}

 \newtheorem{defn}[thm]{Definition}
 
\vspace{20mm}
\begin{document}
\maketitle \vspace{15mm}
\newpage
\begin{abstract}
A new class of  entanglement witnesses (EWs) called reduction
type entanglement witnesses is introduced, which can detect some
multi-qudit entangeled  states including PPT ones with Hilbert
space of dimension $d_{_{1}}\otimes d_{_{2}}\otimes...\otimes
d_{_{n}}$. The novelty of this work comes from the fact that the
feasible regions turn out to be convex polygons, hence the
manipulation of these EWs reduces to linear programming which can
be solved \emph{exactly}  by using simplex method. The
decomposability and non-decomposability of these EWs are studied
and it is shown that it has a close connection with eigenvalues
and optimality of EWs. Also using the Jamio\l kowski isomorphism,
the corresponding possible positive maps,  including the
generalized reduction maps of Ref. \cite{Hall1},  are obtained.

{\bf PACs Index: 03.65.Ud}
\end{abstract}
\vspace{70mm}
\newpage
\section{Introduction}
Quantum entangled states lie at the heart of the rapidly
developing field of quantum information science, which encompasses
important potential applications such as quantum communication,
quantum computation
\cite{Preskill,Nielsen,Ekert}. However, the fundamental nature of
entangled states has tantalized physicists since the earliest days
of quantum mechanics, and even today is by no means fully
understood. One of the most basic problems is that how can one
tell if a quantum state is entangled?, and how entangled is it
still after some noisy quantum process. \\
Here we will deal with the states of finite dimensional Hilbert
space
${\cal{H}}={\cal{H}}_{d_{1}}\otimes...\otimes{\cal{H}}_{d_{n}}$. A
density matrix $\rho$ is called \emph{unentangled} or
\emph{separable}  if it can be written as a convex combination of
pure product states as
$\ket{\gamma}=\ket{\alpha^{(1)}}\otimes...\otimes\ket{\alpha^{(n)}}$.
If no such convex linear combination exists for a given $\rho$, then
the state is called \emph{entangled}. Although, in the case of pure
states of bipartite systems it is to check whether a given state is,
or is not entangled, the question is yet an open problem in the case
of mixed states (see the recent good
reviews\cite{Horod1,Lewen1,Terhal,Bruss}). There have been many
efforts in recent years to analyze the separability and quantitative
character of quantum entanglement. The Bell inequalities satisfied
by a separable system give the first necessary condition for
separability \cite{Bell}. In 1996, Peres made an important step
towards proving that, for a separable state, the partial
transposition with respect to one subsystem of a bipartite density
matrix is positive, $\rho^{T_{A}}\geq0$ \cite{Peres}. By
establishing a close connection between positive map theory and
separability, Horodecki {\it et al}. promptly showed that this is
sufficient condition for separability for bipartite systems of
$2\otimes2$ and $2\otimes3$ \cite{Horod2}. Regarding the
quantitative character of entanglement, Wootters succeeded in giving
an elegant formula to compute the \emph{entanglement of formation}
\cite{Wootters1} of $2\otimes2$ mixtures, thus giving also a
separability criterion \cite{Wootters2}. An alternative method to
detect entanglement is to construct so-called entanglement witnesses
\cite{Horod2,Lewen2,Jafar1,Jafar2} and positive maps \cite{Stromer}.
Entanglement witnesses are physical observables that can
\emph{detect} the presence of entanglement. Recently many attempts
have been made to use the convex optimization methods as a robust
tools in most quantum information areas including construction of
EWs \cite{Doherty1,Doherty2,Doherty3,Jafar1,Jafar2}.\\ The main
motivation for the present investigation is the remarkable fact that
there is no evidence that when and where the manipulation of the EWs
reduces to LP which can be solved exactly. For example, the first
author  and his collaborators have found certain examples of Bell
diagonal EWs and generalized $d \times d$ Choi entanglement
witnesses \cite{Jafar1,Jafar2}, by approximate LP method. Indeed in
most cases determining feasible regions needs to use numerical
calculation and consequently the problem is solved approximately.
Therefore to figure out the problems which can be easily solved
\emph{exactly} in a \emph{optimal} way and also can be generalized
to an \emph{arbitrary} number of particles with different Hilbert
spaces is the main purpose of this paper. To this aim  we introduce
a new class of $(d_{_{1}}\otimes d_{_{2}}\otimes...\otimes
d_{_{n}})$-multi-qudit  EWs called reduction type entanglement
witnesses(REWs) and show that  the computational difficulty in
solving such problems reduces to LP which can be solved by simplex
method\cite{Boyd}. In these cases the feasible regions  are
simplexes and the minimum value of optimization problem is achieved
on their apexes. On the other hand we show that the EWs
corresponding to hyperplanes surrounding feasible regions are
optimal. Another advantage  of this work is the fact that all REWs
can be  written in terms of some positive operators and optimal EWs.
Also in most cases  the decomposability of REWs is rather
determined, where two of their eigenvalues plays an important role
in this issue. Another important consequences of such EWs are the
positive maps including the generalized reduction map \cite{Hall1},
which can be obtained from these EWs or their tensor product,  via
Jamio\l kowski isomorphism. Finally a class of entangled density
matrix including PPTES are provided which can be detected by such
REWs and consequently they determine  the non-decomposability of
REWs. At the end we carry out in detail some particular cases such
as multi-qubit and $2\otimes3\otimes4$ and generalized reduction EW.
\\
The paper organized as follows: In section 2 we give a brief
review of EWs. In section 3 we explain the general scheme of
linear programming. In section 4, we introduce a new class of EWs
which can be put in the realm of LP, since their feasible regions
are convex polygons (indeed simplexes) which can be  exactly
determined.  In section 5, we show that all EWs corresponding to
hyperplanes surrounding  the feasible regions are optimal.
Section 6 is devoted to   some interesting examples such as:
multi-qubit REWs, $(d\otimes d...\otimes d)$ multi-qudit REWs and
$2\otimes3\otimes4$ REW.  Section 7 deals with two examples of LP
type, where  the first one can be solved exactly by the
prescription of this paper. In section 8 we introduce some
entangled and PPT states which can be detected by REWs and the
decomposability of REWs is discussed. In section 9  by using
Jamio\l kowski isomorphism, the relation between REWs and positive
maps is explained. The paper is ended with a brief conclusion and
three appendices.
\section{Entanglement witness}
As  mentioned above in the introduction, one of the pragmatic
approach to detect entanglement is to construct entanglement
witnesses. Let us briefly recall what these operators are.
\begin{defn} A Hermitian operator $\mathcal{W}$ is called an entanglement
witness detecting the entangled state $\rho_{e}$ if \ $
Tr(\mathcal{W}\rho_{e})<0 $ and $ Tr(\mathcal{W}\rho_{s})\geq 0 $
for all separable state $\rho_{s}$.
\end{defn}
So, if we have a state $\rho$ and we measure
$Tr(\mathcal{W}\rho)<0$, we can be sure that $\rho$ is entangled. \\
This definition has a clear geometrical meaning. The expectation
value of an observable depends linearly on the state. Thus, the
set of states where $Tr(\mathcal{W}\rho)=0$ holds is a hyperplane
in the set of all states, cutting this set into two parts. In the
part with $Tr(\mathcal{W}\rho)>0$ lies the set of all separable
states, the other part
( with $Tr(\mathcal{W}\rho)<0$) is the set of state detected by $\mathcal{W}$.
From this geometrical interpretation it follows that all entangled
states can be detected by witness. Indeed for each entangled state
$\rho_{e}$ there exist an entanglement
  witness detecting it \cite{Horod2}.
\begin{defn}
An EW is decomposable (d-EW) iff there exists operators
$\mathcal{P},\ \mathcal{Q}_{i}$ with
\begin{equation}
 \mathcal{W}=\mathcal{P}+\mathcal{Q}_{1}^{T_{A}}+\mathcal{Q}_{2}^{T_{B}}+...+\mathcal{Q}_{N}^{T_{Z}}
 \quad\quad \mathcal{P},\mathcal{Q}_{i}\geq0
\end{equation}
where partial transpose taken with respect to some subsystems  and
it is non-decomposable if it can not be put in this form
\cite{Doherty3}.
\end{defn}
Only non-decomposable EWs can detect PPT entangled states that is
those density matrices which have positive partial transposition
with respect to each subsystems\cite{Lewen3}.
\section{Manipulating EWs by LP method}
This section deals with basic definitions of linear
programming(LP) and general scheme  to construct
EWs by an exact LP method.\\
To this aim  first we consider a Hermitian operator $\mathcal{W}$
with some negative eigenvalues
\begin{equation}
 \mathcal{W}=\sum_{i}a_{i} \sigma_{i}
\end{equation}
where $ \sigma_{i}$ are positive operators, with $0\leq Tr(
\sigma_{i} \rho_{s})\leq1$ for every separable states $
\rho_{s}$. Note that, the condition $0\leq Tr( \sigma_{i}
\rho_{s})\leq1$  is not always required. It is used here  only to
simplify analyzing  the problem and pave the way to generalize
the prescription to  multi-qudits with arbitrary higher dimensions
as will be discussed in the following.\\
The minimum value of $\mathcal{F}=Tr(\mathcal{W}\rho_{s})$
achieves for pure product state, since every mixed $\rho_{s}$ can
be written as a convex combination of pure product states (due to
the convexity of separable region) as $\rho_{s}=\sum_{i}p_{i}
|\gamma_{i}\rangle\langle\gamma_{i}|$ with $p_{i}\geq0$ and
$\sum_{i}p_{i}=1$,  hence we have
\begin{equation}
Tr(\rho_{s}\mathcal{W})=\sum_{i}p_{i}Tr(\mathcal{W}|\gamma_{i}\rangle\langle\gamma_{i}|)\geq
C_{min}
\end{equation}
$$
\mathrm{with}  \quad\quad C_{min}=\min \mathcal{F}:=Tr(\mathcal{W}
|\gamma\rangle\langle\gamma|)\quad |\gamma\rangle \in D_{prod},$$
where $D_{prod}$ denotes the set of product states. Thus  we need
to  find the pure product state $|\gamma_{min}\rangle$ which
minimize $Tr(\mathcal{W}|\gamma\rangle\langle\gamma|)$. Now, as
the pure product state $|\gamma\rangle$ varies, the map defined by
$P_{i}=Tr(\sigma_{i}|\gamma\rangle\langle\gamma|)$ maps the set
$D_{prod}$ into a region  inside the hypercube defined by $0\leq
P_{i}\leq 1\quad i=1,2,...,N$. This is not all circumstances. An
important and difficult task is to find the convex region (called
feasible region) inside
 this hypercube which comes from
$\mathcal{F}_{_{\mathcal{W}}}:=Tr(\mathcal{W}\rho_{s})$ as
$\rho_{s}$ varies on $D_{sep}$, where $D_{sep}$ denotes the set of
separable states. Here in this work we are interested in the EWs
with the feasible regions of  simplexes (or at most convex
polygons)types, such that the manipulating these EWs  amounts  to
\begin{equation}
 \mathrm{minimize}\quad \mathcal{F}_{_{\mathcal{W}}}:=Tr(\mathcal{W}\rho_{s})=\sum_{i}a_{i}P_{i}
\end{equation}
$$
 \mathrm{subject \ to}\quad \sum_{i=1}^N(c_{ij}P_{i}-d_{i})\geq 0 \quad  j=1,2,...
$$
where $c_{ij}\; , d_{i} \; i,j=1,2,... $are parameters of
hyper-planes surrounding  the feasible regions. Therefore, the
corresponding boundary points of  feasible region will minimize
exactly $\mathcal{F}_{_{\mathcal{W}}}$, thus  the required task
reduces to LP problem which can be solved  by the well-known
simplex method \cite{Boyd}.
\section{$(d_{1}\otimes d_{2}\otimes...\otimes d_{n})$multi-qudit reduction type EWs}
In this section we consider $n$ Particles with arbitrary
dimensions. Without loss of generality particles can be arranged
so that $d_{1}\leq
 d_{2}\leq...\leq d_{n}$. The
discussion of  some special cases is postponed to  Section
\ref{special}.
 We introduce and parameterize the multi-qudit reduction-type
entangled witnesses (REWs) labeled by subscript R as
\begin{equation}\label{wit}
\mathcal{W}_{_{R}}^{(n)}=\sum_{ S\subsetneqq N'}b_{_{S}}\
\sigma_{_{S}}+d_{1}\;
b_{_{2,...,n}}|\psi_{_{00...0}}\rangle\langle\psi_{_{00...0}}|
 +\sum_{S\subsetneqq N'}a'_{_{S}} \sigma'_{_{S}},
\end{equation}
 where  $N'=\{2,...,n\}$ , with
 $b_{_{\emptyset}}=b_{_{1}}$ , $a'_{_{\emptyset}}=a'_{_{1}}$
 and  $\sigma_{_{S}}$ , $\sigma'_{_{S}}$
defined as
\begin{equation}
 \sigma_{_{S}}=\sum_{i=0}^{d_{1}-1}|i\rangle\langle i|\otimes O_{i}^{(2)}\otimes...\otimes O_{i}^{(n)}
 \quad\quad \mathrm{with}\quad\quad
 O_{i}^{(k)}= \left\{\begin{array}{c}
  |i\rangle\langle i| \quad \mathrm{if}\quad k\in S \\
  I   \quad\quad   \mathrm{if}\quad k\notin  S\\
\end{array}\right.
\end{equation}
  and
\begin{equation}\label{rho'}
\sigma'_{_{S}}=\sum_{i_{1}=0}^{d_{1}-1}\sum_{
i_{2}=0}^{d_{2}-1}...\sum_{ i_{n}=0}^{d_{n}-1}
|i_{1}\rangle\langle i_{1}|\otimes
O_{i_{2}}'^{(2)}\otimes...\otimes O_{i_{n}}'^{(n)}\hspace{3cm}
\end{equation}
with
$$O_{i_{k}}'^{(k)}
 = \left\{\begin{array}{c}
  |i_{1}\rangle\langle i_{1}|\delta_{i_{1}i_{k}}\hspace{3cm}\mathrm{if}\quad k\in S \hspace{7cm}\\
  |i_{k}\rangle\langle i_{k}|\prod_{k\neq m=1}^{n}(1-\delta_{i_{k}i_{m}})
  \quad  \mathrm{if}\quad k\notin S,\quad |S|\leq n-3 \quad\mathrm{and}\quad i_{2},...,i_{n}\leq d_{1}-1\\
   0\quad\quad   \hspace{4cm}\mathrm{if}\quad k\notin S ,\quad |S|=n-2\quad\mathrm{and}\quad i_{2},...,i_{n}\leq d_{1}-1\\
 |i_{k}\rangle\langle i_{k}| (1-\delta_{i_{k}i_{1}})\hspace{2cm}\mathrm{otherwise}\hspace{7cm}\\
\end{array}\right.
$$
 , respectively and
\begin{equation}\label{maxs}
|\psi_{_{00...0}}\rangle:=\frac{1}{\sqrt{d_{1}}}\sum_{i=0}^{d_{1}-1}|i^{(1)}\rangle|i^{(2)}\rangle...|i^{(n)}\rangle
\end{equation}
is maximally  entangled state.
 In this notation we have
$\sigma'_{_{\emptyset}}=\sigma'_{1}$. Obviously for multi-qubit
system none of $\sigma'_{_{S}}$ exists. The number of $P'_{S}$
depends on the dimensions of particles $d_{i}$'s and it can take one
of the following values
$$
m=\left\{\begin{array}{c}
 \sum_{|S|=n-d}^{n-3}\mathbf{C}_{|S|}^{n-1}\quad \mathrm{if }\quad
d_{1}=d_{2}=...=d_{n}=d \\
  \sum_{|S|=0}^{n-2}\mathbf{C}_{|S|}^{n-1}-(2^{m-1}-1)\quad
\mathrm{if }\quad d_{1}=...=d_{m}<d_{m+1}\leq
...\leq d_{n} \\
\end{array}\right.
$$
 We introduce the new  more convenient
parameters  $ a_{_{1}}=b_{_{1}}=b_{\emptyset}$ ,
$a_{_{S}}=b_{_{S}}+\sum_{_{ S'\subsetneqq S}}b_{_{S'}}$, instead
of $b_{_{S}}$'s. In order to turn the   observable (\ref{wit}) to
an EW, we need to choose its parameters in such a way that it
becomes a non-positive operator with positive  expectation values
in any pure product state
$\ket{\gamma}=\ket{\alpha^{(1)}}\otimes...\otimes\ket{\alpha^{(n)}}.$
\\
Now it is the time to reduce the problem to LP one. Io order to
determine the feasible region,  we need to know the apexes,
namely the extremum points, to construct the hyperplanes
surrounding the feasible regions. Suppose that $\ket{\gamma}$ be a
pure  product state with $
|\alpha^{(k)}\rangle=(\alpha_{_{0}}^{(k)},\alpha_{_{1}}^{(k)},...,\alpha_{_{d_{_{k}}-1}}^{(k)})^{T}$
and let
$$
P_{S}:=Tr(\sigma_{_{S}}\ket{\gamma}\bra{\gamma})=\sum_{i=0}^{d_{1}-1}|\beta_{_{i}}^{(1)}\beta_{_{i}}^{(2)}...
\beta_{_{i}}^{(n)}|^2 \quad,\quad\quad \beta_{_{i}}^{(k)}
 = \left\{\begin{array}{c}
  \alpha_{_{i}}^{(k)}\quad \mathrm{if}\quad k\in \{1\}\cup S \\
  1  \quad\quad   \mathrm{if}\quad k\notin \{1\}\cup S\\
\end{array}\right.
$$

$$
P_{2,...,n}:=d_{1}\
|\langle\psi_{_{00...0}}\ket{\gamma}|^{2}=\left|\sum_{i=0}^{d_{1}-1}\alpha_{_{i}}^{(1)}\alpha_{_{i}}^{(2)}...
\alpha_{_{i}}^{(n)}\right|^2
$$
$$
P'_{S}:=Tr(\sigma'_{_{S}} \ket{\gamma}\bra{\gamma}),
$$
which all lie in the interval $[0,1]$ (see appendix I). The number
of $P_{S}$ is $2^{n-1}-1$ (cardinality of the power set of $N'$
excepted $\{\emptyset\}$). \\   The extremum points or apexes consist of\\
\textbf{a.} origin: $\quad P_{S}=0,\quad  P'_{S}=0\; \forall
S\subseteqq N'
 \;\; $ which  corresponds to the following choice of pure product state
$$|\alpha^{(1)}\rangle=(1\ \ 0\ 0 \ ...\ 0)^T$$
$$|\alpha^{(k)}\rangle=(0\ 1\ 0\ ...\ 0)^T \quad  k\in N'$$
\textbf{b.} $\quad P_{S}=1,\quad P_{S'}=1\;\forall S'\subseteqq
S,\quad\quad P_{N'}=0,\quad P_{S''}=0 \;\forall S''\subseteqq
N'\setminus S,\quad P'_{S}=0\; \forall S\subseteqq N'$\\which can be
reached by choosing the following pure product states
$$|\alpha^{(k)}\rangle=(1\ \ 0\ 0 \ ...\ 0)^T \quad  k\in\{1\} \cup S,$$
$$|\alpha^{(k)}\rangle=(0\ 1\ 0\ ...\ 0)^T \quad  k\notin \{1\}\cup S.$$
Obviously if $P_{S}=1$, then for all  $S'\subseteqq S$, $P_{S'}=1,$
thus for   $S=N'$, we get the following important apex  \\
$\textbf{c.} \quad P_{S}=1,\; P'_{S}=0\; \forall
S\subseteqq N'.\hspace{10.2cm}$\\
which can be obtained  by choosing
$$|\alpha^{(k)}\rangle=(1\ \ 0\ 0 \ ...\ 0)^T \quad \forall k\in\{1\} \cup N'.$$
\\
$\textbf{d.}\quad   P'_{S}=1,\;
 P_{S'}=1\; \forall S'\subseteqq S\subseteqq N'\quad, \mathrm{the \ others\ are \ zero.}$\\
The last category  arise from pure product state with the components
of the form $\alpha_{0}^{(1)}=\alpha_{0}^{(k)}=1$, if $ k\in S$ and
$\alpha_{k-1}^{(k)}=1 $, if $ k\notin S$ , i.e.,
$$|\alpha^{(k)}\rangle=(1\ \ 0\ 0 \ ...\ 0)^T \quad  k\in \{1\} \cup S,$$
$$|\alpha^{(k)}\rangle=(0\ \ 0 \ ...\ 0\ \underbrace{1}_{k-1'th}\ 0\ ...\ 0)^T \quad  k\notin  \{1\}\cup S.$$
Regarding the above consideration, we are now ready to state the
feasible region.\\
To this aim we first prove that, $2^{n-1}+m$ extremum points
obtained above form the apexes of $(2^{n-1}+m)$-simplex in Euclidean
space of dimension $N=2^{n-1}+m-1$. To this purpose we consider the
convex hull of theses points, i.e. draw $N+1$ hyperplanes passing
through each combination of  $N$ points out of $N+1$ ones(
$\mathbf{C}_{N}^{N+1}=N+1$). Now, we get a bounded region formed
from their intersection which is the required feasible region of $
\mathcal{W}_{_{R}}^{(n)}$ and it   is obviously a $(2^{n-1}+m)$-
simplex. It is strait forward to show that the feasible region  can
be obtained by taking the expectation values of
\begin{equation}
    ^{(S)}{\mathcal{W}_{_{opt}}^{(n)}}=a_{_{S}}\left(\sigma_{_{S}}
   +\sum_{_{ S\subsetneqq
  S'\neq N'}}(-1)^{|S|+|S'|}\sigma_{_{S'}}+d_{1}\;
(-1)^{|S|+|N'|}|\psi_{_{00...0}}\rangle\langle\psi_{_{00...0}}|-\sigma'_{_{S}}\right)
\quad S\subsetneqq N'
\end{equation}
in  pure product states (see appendix II). Now, in order that
$\mathcal{W}_{_{R}}^{(n)}$ to be  an EW, the following
expectation values
\begin{equation}\label{cordinate}
 \mathcal{F}(P_{2},P_{3},...,P'_{1},P'_{2},...):=Tr(\ \mathcal{W}_{_{R}}^{(n)}\ket{\gamma}\bra{\gamma}\
)=\sum_{ S\subsetneqq N'}b_{_{S}}\ P_{S}+
b_{_{2,...,n}}P_{2,...,n}+\sum_{ S\subsetneqq N'}a'_{_{S}}\ P'_{S}
\end{equation}
must  be  positive. So our task is  to solve the following LP
problem
$$
\mathrm{minimize} \quad\;\; \sum_{ S\subsetneqq N'}b_{_{S}}\
P_{S}+ b_{_{2,...,n}}P_{2,...,n}+\sum_{ S\subsetneqq N'}a'_{_{S}}\
P'_{S} \vspace{-3mm}
$$
\begin{equation}\label{region}
 \mathrm{subject\ to} \quad\;\;\left\{\begin{array}{c}
  P_{S}- P'_{S}+\sum_{_{ S\subsetneqq
  S'}}(-1)^{|S|+|S'|}P_{S'}\geq0 \\
 \forall\ P'_{S}\geq0 \hspace{5cm} \\
\end{array}\right.
\end{equation}
 Putting the coordinates of apexes in Eq.(\ref{cordinate})
 yields all $a_{_{S}}\geq0$ and $a_{_{S}}+a'_{_{S}}\geq0$.
 As we stated in  previous section, all $P_{S}$ and $P'_{S}$ lie in the
 closed  interval $[0,1]$.
Now, $\mathcal{F} $ is a linear function of $P_{S}$ and $P'_{S}$
and if  we require it to be positive on the apexes ( which are
extremum points), then it will be positive in the whole feasible
region.\\
 At the end  we  need to know all
eigenvalues of $\mathcal{W}_{_{R}}^{(n)}$ which consist of
$a_{_{S}},a_{_{S}}+a'_{_{S}}, \omega_{_{1}}=a_{_{N'}}-b_{_{N'}}\
\mathrm{and} \
\omega_{_{2}}=d_{1}a_{_{N'}}-(d_{1}-1)\omega_{_{1}}$. Since
$a_{_{S}},a_{_{S}}+a'_{_{S}}\geq0$, then one of the remaining
eigenvalues: $\omega_{_{1}}$ and $\omega_{_{2}}$ must be negative
 to  guarantee   $\mathcal{W}_{_{R}}^{(n)}$ to be  an EW.
\section{Optimality of
$^{(S)}\mathcal{W}_{_{opt}}^{(n)}$}\label{optimal} After
determining the feasible regions, one needs to know whether  the
boundary of EWs is  formed by optimal EWS.  An EW is optimal  if
a positive operator $P$ is subtracted from
that then it will be no longer an EW \cite{Lewen2}.\\
Note that the EWs corresponding to hyperplanes surrounding
feasible regions of $\mathcal{W}_{_{R}}^{(n)}$  are optimal since
they cover the simplex feasible region in an optimal way (see
appendix III). Thus, the structure of the optimal EWs
$^{(S)}\mathcal{W}_{_{opt}}^{(n)}$ characterizes the boundary of
REWs $\mathcal{W}_{_{R}}^{(n)}$. In fact, from the results of this
Section it will become clear that we can restrict ourselves to the
structure of the optimal EWs corresponding to hyperplanes
surrounding feasible regions. In other words optimal EWs
$^{(S)}{\mathcal{W}_{_{opt}}^{(n)}}$
are tangent to the boundary  between separable and non-separable states.\\
Another advantage of $^{(S)}{\mathcal{W}_{_{opt}}^{(n)}}$ is that
one can rewrite the $\mathcal{W}_{_{R}}^{(n)}$ in terms of
positive operators
$\sigma'_{_{S}},|\psi_{_{00...0}}\rangle\langle\psi_{_{00...0}}|$
and some optimal EWs, i.e.,
\begin{equation}\label{decom}
\mathcal{W}_{_{R}}^{(n)}=\sum_{S}a_{_{S}}{^{(S)}\mathcal{W}_{_{opt}}^{(n)}}
+d_{_{1}}\omega_{_{2}}|\psi_{_{00...0}}\rangle\langle\psi_{_{00...0}}|+\sum_{S}(a_{_{S}}+a'_{_{S}})\sigma'_{_{S}}\quad\quad
S\subsetneqq N'
\end{equation}
Therefore for positive  $\omega_{_{2}}$, the REWs can be
decomposed as
\begin{equation}
\mathcal{W}_{_{R}}^{(n)}=\sum_{S}{^{(S)}\mathcal{Q}}^{T_{(N'\setminus
S)}}+\mathcal{P}
\end{equation}
 where $^{(S)}\mathcal{Q}:={^{(S)}\mathcal{W}_{_{opt}}^{(n)}}^{T_{(N'\setminus S)}}$ and $\mathcal{P}$
is
 positive operator and in this case $\mathcal{W}_{_{R}}^{(n)}$ can not
 detect PPT entangled states( non-separable  density matrices with positive partial
 transpose with respect to all particles).
\section{Some special cases of $\mathcal{W}_{_{R}}^{(n)}$}\label{special}
In this section we discuss some special cases of REWs  to enlighten
the subject.
\subsection{Multi-qubit reduction type EWs}
It is important both theoretically and experimentally to study
multi-qubit entanglement and to provide EWs to verify that in a
given multi-qubit state, entanglement is really present. The Eq.
(\ref{wit}) for system of n-qubits reduces to
\begin{equation}
\label{witness} \mathcal{W}_{_{R}}^{(n)}=\sum_{S\subsetneqq
N'}b_{_{S}}\ \sigma_{_{S}}+2\;
b_{_{2,...,n}}|\psi_{_{00...0}}\rangle\langle\psi_{_{00...0}}|
\end{equation}
As mentioned before, the  dimension of qubit-systems does not allow
presence of $\sigma'_{_{S}}$.

The number of P's is $2^{n-1}-1\ $, whereas the number of apexes is
$2^{n-1}$.
 Putting the coordinates of apexes in Eq.(\ref{cordinate})
again indicates that all $a_{_{S}}\geq0$.  The feasible region is
$2^{n-1}$-simplex  of dimension $2^{n-1}-1\ $ surrounded by
hypersurfaces defined by (\ref{region}) with all $P'$'s eliminated.

\subsubsection{Three-qubit (n=3)}
The first nontrivial example of REWs for multi-qubit system is
three qubit REW
\begin{equation}\label{three-EW}
\mathcal{W}_{_{R}}^{(3)}=a_{_{1}}I_{_{8}}+
(a_{_{2}}-a_{_{1}})\sigma_{_{2}}+(a_{_{3}}-a_{_{1}})\sigma_{_{3}}+2\
(a_{_{2,3}}+a_{_{1}}-a_{_{2}}-a_{_{3}})|\psi_{_{000}}\rangle\langle\psi_{_{000}}|
\end{equation}
with eigenvalues $a_{_{1}},a_{_{2}},a_{_{3}}\geq0 ,
\omega_{_{1}}=a_{_{2}}+a_{_{3}}-a_{_{1}}$ and $\omega_{_{2}}= 2\
a_{_{2,3}}-\omega_{_{1}} $.
 In order to  obtain the feasible region,  we need the
coordinates of apexes which are
$$
\left\{\begin{array}{c}
  \ket{\gamma}=\left(\begin{array}{c}
               1 \\
               0 \\
             \end{array}\right)\otimes\left(\begin{array}{c}
               0\\
               1 \\
             \end{array}\right)\otimes\left(\begin{array}{c}
              0 \\
              1 \\
             \end{array}\right)\longrightarrow(P_{2}=0,P_{3}=0,P_{2,3}=0) \\
  \ket{\gamma}=\left(\begin{array}{c}
               1\\
               0 \\
             \end{array}\right)\otimes\left(\begin{array}{c}
                1 \\
               0 \\
             \end{array}\right)\otimes\left(\begin{array}{c}
               0 \\
               1\\
             \end{array}\right)\longrightarrow(P_{2}=1,P_{3}=0,P_{2,3}=0) \\
  \ket{\gamma}=\left(\begin{array}{c}
               1 \\
               0 \\
             \end{array}\right)\otimes\left(\begin{array}{c}
                0 \\
               1 \\
             \end{array}\right)\otimes\left(\begin{array}{c}
               1 \\
               0 \\
             \end{array}\right)\longrightarrow(P_{2}=0,P_{3}=1,P_{2,3}=0) \\
            \ket{\gamma}=\left(\begin{array}{c}
               1 \\
               0 \\
             \end{array}\right)\otimes\left(\begin{array}{c}
                1 \\
               0 \\
             \end{array}\right)\otimes\left(\begin{array}{c}
               1 \\
               0 \\
             \end{array}\right)\longrightarrow(P_{2}=1,P_{3}=1,P_{2,3}=1) \\
\end{array}\right.
$$
As we see the number of $P_{S}$ is three, while the number of
optimal points is four. Therefore, there are four hyperplanes
surrounding the feasible region, where the hyperplanes pass
through each combination  of   three points out of four ones.
(see Fig.1). Thus the problem can be reduced to
$$
\hspace{-3cm}\mathrm{minimize}\quad\quad\quad\quad
Tr(\mathcal{W}_{_{R}}^{(3)}|\gamma\rangle\langle\gamma|)
$$
\begin{equation}
 \mathrm{subject\ to} \quad\;\;\left\{\begin{array}{c}
1+P_{2,3}-P_{2}-P_{3}\geq0,\\
 P_{2}-P_{2,3}\geq0,\\
 P_{3}-P_{2,3}\geq0,\\
P_{2,3}\geq0,\\
\end{array}\hspace{2cm}\right.
\end{equation}
where the above given inequalities follow rather easily by taking
the expectation value of the following optimal EWs
\begin{equation}
\begin{array}{c}
^{2}\mathcal{W}_{opt}^{(3)}= a_{_{2}}(\ \sigma_{_{2}}-2\
|\psi_{_{000}}\rangle\langle\psi_{_{000}}|) \\
^{3}\mathcal{W}_{opt}^{(3)}= a_{_{3}}(\ \sigma_{_{3}}-2\
|\psi_{_{000}}\rangle\langle\psi_{_{000}}|) \\
^{1}\mathcal{W}_{opt}^{(3)}= a_{_{1}}(I_{_{8}}-
\sigma_{_{2}}-\sigma_{_{3}}+2\
|\psi_{_{000}}\rangle\langle\psi_{_{000}}|)\\
\end{array}
\end{equation}
 in pure product states, and as usual   the  optimal EWs can be written as partial transpose of
 the following positive operators
\begin{equation}
\begin{array}{c}
^{1}{\mathcal{W}_{opt}^{(3)}}^{T_{23}}=a_{_{1}} (|100\rangle+|011\rangle)(\langle100|+\langle011|),\\
^{2}{\mathcal{W}_{opt}^{(3)}}^{T_{3}}=
a_{_{2}}(|001\rangle+|110\rangle)(\langle001|+\langle110|), \\
^{3}{\mathcal{W}_{opt}^{(3)}}^{T_{2}}=
a_{_{3}}(|010\rangle+|101\rangle)(\langle010|+\langle101|),\\
\end{array}
\end{equation}
respectively. Now the EW  $\mathcal{W}_{_{R}}^{(3)}$ can be written
in terms of $^{i}\mathcal{W}_{opt}$ as
\begin{equation}
\begin{array}{c}
  \mathcal{W}_{_{R}}^{(3)}=a_{_{1}}{^{1}\mathcal{W}_{opt}}+
a_{_{2}}{^{2}\mathcal{W}_{opt}}+a_{_{3}}{^{3}\mathcal{W}_{opt}}+2\
\omega_{_{2}}|\psi_{_{000}}\rangle\langle\psi_{_{000}}| \\
  \;\;\;=a_{_{1}}{^{(1)}\mathcal{Q}}^{T_{(23)}}+ a_{_{2}}{^{(2)}\mathcal{Q}}^{T_{(3)}}
  +a_{_{3}}{^{(3)}\mathcal{Q}}^{T_{(2)}}+2\
\omega_{_{2}}\mathcal{P} \\
\end{array}
\end{equation}
As mentioned above in section $4$, for $ \omega_{_{2}}\geq$ the EW
$\mathcal{W}_{_{R}}^{(3)}$ becomes decomposable and can not detect
PPT entangled states.
\subsection{n-qudit $(d\otimes d\otimes ...\otimes d)$}
For n particles with the same dimensions the extra terms
$\sigma'_{_{S}}$ will appear in EWs, provided that  the requirement
$|S|\leq n-3$ is met. Then the Eq.(\ref{wit}) becomes
\begin{equation}
\mathcal{W}_{_{R}}^{(n)}=\sum_{ S\subsetneqq N'}b_{_{S}}\
\sigma_{_{S}}+d\;
b_{_{2,...,n}}|\psi_{_{00...0}}\rangle\langle\psi_{_{00...0}}|
 +\sum_{S\subsetneqq N',|S|\leq n-3}a'_{_{S}} \sigma'_{_{S}}
\end{equation}
where
$$
\begin{array}{cc}
 \sigma'_{_{S}}=\sum_{i_{1},...,i_{n}=0}^{d-1} |i_{1}\rangle\langle i_{1}|\otimes
O_{i_{2}}'^{(2)}\otimes...\otimes O_{i_{n}}'^{(n)}, & \quad
O_{i_{k}}'^{(k)}
 = \left\{\begin{array}{c}
  |i_{1}\rangle\langle i_{1}|\delta_{i_{1}i_{k}}\hspace{3cm}\quad k\in S \hspace{7cm}\\
  |i_{k}\rangle\langle i_{k}|\prod_{k\neq m=1}^{n}(1-\delta_{i_{k}i_{m}})
  \quad \quad k\notin S.\hspace{7cm} \\
\end{array}\right. \\
\end{array}
$$
We discuss below the most simple case: an REW consisting of just
two qudits with the same dimension  $d$, that is
$$
\mathcal{W}_{_{R}}^{(2)}=a_{_{1}}I_{_{d^2}}+d\
(a_{_{2}}-a_{_{1}})|\psi_{_{00}}\rangle\langle\psi_{_{00}}|,
$$
Where for $a_{_{2}}=0$ it reduces to the  well known reduction EW
( the term  "reduction-type EWs"  for general
$\mathcal{W}_{_{R}}^{(n)}$ is  inspired  from this particular
case). In this case we have only $P_{2}$  which can takes values
between 0 and 1. So the feasible region is just the line segment
$0\leq P_{2}\leq1$. Putting 0 and 1 in
$\mathcal{F}(P_{2})=a_{_{1}}+(a_{_{2}}-a_{_{1}})P_{2}\ $ yields
$a_{_{1}}\geq0$ and $a_{_{2}}\geq0$ respectively. The eigenvalues
are $ a_{_{1}}$ and $2\ a_{_{2}}-a_{_{1}}$ where the second  one
must be negative to ensure detecting some entangled states.
\subsection {Three particles $(2\otimes 3\otimes 4)$}
As a particular example of REWs with different dimension let us
discuss three particles with $2\otimes 3\otimes4$ dimensions
\begin{equation}
\mathcal{W}_{_{R}}^{(3)}=a_{_{1}}I_{_{8}}+
(a_{_{2}}-a_{_{1}})\sigma_{_{2}}+(a_{_{3}}-a_{_{1}})\sigma_{_{3}}+2\
(a_{_{2,3}}+a_{_{1}}-a_{_{2}}-a_{_{3}})|\psi_{_{000}}\rangle\langle\psi_{_{000}}|+
a'_{_{1}}\sigma'_{_{1}}+a'_{_{2}}\sigma'_{_{2}}+a'_{_{3}}\sigma'_{_{3}}
\end{equation}
where we have
$$
\sigma'_{_{1}}=|012\rangle\langle012|+|013\rangle\langle013|+|021\rangle\langle021|+|022\rangle\langle022|
+|023\rangle\langle023|
$$
$$
|102\rangle\langle102|+|103\rangle\langle103|+|120\rangle\langle120|+|122\rangle\langle122|+|123\rangle\langle123|
$$
$$
\sigma'_{_{2}}=|002\rangle\langle002|+|003\rangle\langle003|+|112\rangle\langle112|+|113\rangle\langle113|
$$
$$
\sigma'_{_{3}}=|020\rangle\langle020|+|121\rangle\langle121|
$$
Here all possible $\sigma'_{S} \; ( S=\{2\} , \{3\}\ \mathrm{and}\ \emptyset )$  can be appear.
 In this case feasible region lies in a space of dimension
six. The coordinates of apexes and relevant pure product states
$|\gamma\rangle$  are
$$
\hspace{7.4cm} (P_{2},P_{3},P_{23},P'_{1},P'_{2},P'_{3})
$$
$$
\begin{array}{cc}
    |\gamma\rangle=\left(\begin{array}{c}
    1 \\
    0 \\
  \end{array}\right)\otimes\left(\begin{array}{c}
               0 \\
               1 \\
               0 \\
             \end{array}\right)\otimes\left(\begin{array}{c}
                          0 \\
                          1 \\
                          0 \\
                          0 \\
                        \end{array}\right)\longrightarrow
   & (0,0,0,0,0,0) \\
\end{array}
$$
$$
\begin{array}{cc}
  |\gamma\rangle=\left(\begin{array}{c}
    1 \\
    0 \\
  \end{array}\right)\otimes\left(\begin{array}{c}
               1 \\
               0\\
               0 \\
             \end{array}\right)\otimes\left(\begin{array}{c}
                          0 \\
                          1 \\
                          0 \\
                          0 \\
                        \end{array}\right)\longrightarrow
  & (1,0,0,0,0,0) \\
\end{array}
$$
$$
 \begin{array}{cc}
  |\gamma\rangle=\left(\begin{array}{c}
    1 \\
    0 \\
  \end{array}\right)\otimes\left(\begin{array}{c}
               0 \\
               1 \\
               0 \\
             \end{array}\right)\otimes\left(\begin{array}{c}
                          1\\
                          0 \\
                          0 \\
                          0 \\
                        \end{array}\right)\longrightarrow
  & (0,1,0,0,0,0) \\
\end{array}
$$
$$
\begin{array}{cc}
  |\gamma\rangle=\left(\begin{array}{c}
    1 \\
    0 \\
  \end{array}\right)\otimes\left(\begin{array}{c}
               1 \\
               0 \\
               0 \\
             \end{array}\right)\otimes\left(\begin{array}{c}
                          1 \\
                          0 \\
                          0 \\
                          0 \\
                        \end{array}\right)\longrightarrow
  & (1,1,1,0,0,0) \\
\end{array}
 $$
 $$
\begin{array}{cc}
  |\gamma\rangle=\left(\begin{array}{c}
    1 \\
    0 \\
  \end{array}\right)\otimes\left(\begin{array}{c}
               0 \\
               1 \\
               0 \\
             \end{array}\right)\otimes\left(\begin{array}{c}
                          0 \\
                          0 \\
                          1 \\
                          0 \\
                        \end{array}\right)\longrightarrow
  & (0,0,0,1,0,0) \\
\end{array}
  $$
 $$
\begin{array}{cc}
  |\gamma\rangle=\left(\begin{array}{c}
    1 \\
    0 \\
  \end{array}\right)\otimes\left(\begin{array}{c}
               1 \\
               0 \\
               0 \\
             \end{array}\right)\otimes\left(\begin{array}{c}
                          0 \\
                          0 \\
                          1 \\
                          0 \\
                        \end{array}\right)\longrightarrow
  & (1,0,0,0,1,0) \\
\end{array}
  $$
 $$
\begin{array}{cc}
  |\gamma\rangle=\left(\begin{array}{c}
    1 \\
    0 \\
  \end{array}\right)\otimes\left(\begin{array}{c}
               0 \\
               1 \\
               0 \\
             \end{array}\right)\otimes\left(\begin{array}{c}
                          1 \\
                          0 \\
                          0 \\
                          0 \\
                        \end{array}\right)\longrightarrow
  & (0,1,0,0,0,1) \\
\end{array}
$$
Again choosing all combinations of six apexes out of  seven ones,
one can find the boundary of feasible region as
$$
\begin{array}{c}
P'_{1},P'_{2},P'_{3},P_{2,3}\geq0\\
 P_{2}-P_{2,3}-P'_{2}\geq0\\
 P_{3}-P_{2,3}-P'_{3}\geq0\\
1+P_{2,3}-P_{2}-P_{3}-P'_{1}\geq0\\
\end{array}
$$
where the EWs corresponding  to these hyperplanes are

\begin{equation}
\begin{array}{c}
^{2}\mathcal{W}_{opt}= a_{_{2}}(\ \sigma_{_{2}}-2\
|\psi_{_{000}}\rangle\langle\psi_{_{000}}|-\sigma'_{2}), \\
^{3}\mathcal{W}_{opt}= a_{_{3}}(\ \sigma_{_{3}}-2\
|\psi_{_{000}}\rangle\langle\psi_{_{000}}|-\sigma'_{3}), \\
^{1}\mathcal{W}_{opt}= a_{_{1}}(I_{_{24}}-
\sigma_{_{2}}-\sigma_{_{3}}+2\
|\psi_{_{000}}\rangle\langle\psi_{_{000}}|-\sigma'_{1}).\\
\end{array}
\end{equation}

Taking the partial transposition of $^{i}\mathcal{W}_{opt}$,
$i=1,2,3$ with respect to $\{2,3\}\setminus\{i\}$ yields
\begin{equation}
\begin{array}{c}
^{1}\mathcal{W}_{opt}^{T_{23}}=a_{_{1}} (|100\rangle+|011\rangle)(\langle100|+\langle011|),\\
^{2}\mathcal{W}_{opt}^{T_{3}}=
a_{_{2}}(|001\rangle+|110\rangle)(\langle001|+\langle110|), \\
^{3}\mathcal{W}_{opt}^{T_{2}}=
a_{_{3}}(|010\rangle+|101\rangle)(\langle010|+\langle101|),\\
\end{array}
\end{equation}
respectively. Evidently these EWs are optimal, since these  have
been written as the partial transposition of pure maximally
entangled states.
\section{Bell-diagonal EWs by LP methods}
Recently multi-qubit Bell decomposable entangled witnesses (BDEWs)
\cite{Jafar1} have been introduced as
\begin{equation}
\mathcal{W}_{_{BD}}=\sum_{_{i_{1}i_{2}...i_{n}=0,1}}
a_{_{i_{1}i_{2}...i_{n}}}
\ket{\psi_{_{i_{1}i_{2}...i_{n}}}}\bra{\psi_{_{i_{1}i_{2}...i_{n}}}}
\end{equation}
 where $\ket{\psi_{_{i_{1}i_{2}...i_{n}}}}$ ($d_{i}=2 , i=1,2...,n)$  are n-qubit  maximally entangled orthonormal
states, i.e.,
\begin{equation}
\ket{\psi_{_{i_{1}i_{2}...i_{n}}}}=(\sigma_{z})^{i_{1}}\otimes
(\sigma_{x})^{i_{2}}\otimes ... \otimes (\sigma_{x})^{i_{n}}\ket{\psi_{_{00...0}}}
\end{equation}
 where $\sigma_{x}$ and $\sigma_{z}$ are
usual Pauli matrices.

In general it is hard to  find the BDEWs with feasible region of
simplex type or even polygon type, namely those which can
manipulated by LP method.  Here we give two examples which are
both set in LP problem,  where only one of them (the first
example) can be solved exactly by the prescription of this paper.

The first example is EW of the form
\begin{equation}\label{W1}
\mathcal{W}_{1}=aI_{2^{n}}+2(b-a)\ket{\psi_{
 _{00...0}}}\bra{\psi_{ _{00... 0}}}+2(c-a)\ket{\psi_{
_{00...01}}}\bra{\psi_{ _{00...01}}}+(d-a) \sigma
\end{equation}
 where
$$
\sigma=I_{2^n}-\{(|\psi_{_{011...110}}\rangle\langle
\psi_{_{011...110}}|+|\psi_{_{11...10}}\rangle\langle
\psi_{_{11...10}}|)+(|\psi_{_{011...1}}\rangle\langle
\psi_{_{011...1}}|+|\psi_{_{11...11}}\rangle\langle
\psi_{_{11...11}}|)$$
$$\quad\quad +(|\psi_{_{00...0}}\rangle\langle
\psi_{_{00...0}}|+|\psi_{_{10...0}}\rangle\langle
\psi_{_{10...0}}|)+(|\psi_{_{00...01}}\rangle\langle
\psi_{_{00...01}}|+|\psi_{_{10...01}}\rangle\langle
\psi_{_{10...01}}|)\}
$$
The eigenvalues of $ \mathcal{W}_{1}$ are $a,2b-a,2c-a,d$. This BDEW
is similar to the one  introduced in \cite{Jafar1}, where the extra
term $\sigma$ is added to optimize the EWs corresponding to the
boundary plane of  feasible region. Suppose that $ P_{_{00...0}}=2\
|\langle\psi_{_{00...0}}\ket{\gamma}|^{2}$ , $
  P_{_{00...01}}=2\
|\langle\psi_{_{00...01}}\ket{\gamma}|^{2}$ and $
  P=Tr(\sigma |\gamma\rangle\langle\gamma|)$.
Then the pure  product states which produce the apexes are
$$
\left\{\begin{array}{c}
  \ket{\gamma}=\left(\begin{array}{c}
               1 \\
               0 \\
             \end{array}\right)\otimes\left(\begin{array}{c}
               0 \\
               1 \\
             \end{array}\right)\otimes...\otimes\left(\begin{array}{c}
               0\\
              1 \\
             \end{array}\right)\longrightarrow(P_{_{00...0}}=0,P_{_{00...01}}=0,P=0) \\
  \ket{\gamma}=\left(\begin{array}{c}
               1\\
               0 \\
             \end{array}\right)\otimes\left(\begin{array}{c}
                0 \\
               1 \\
             \end{array}\right)\otimes\left(\begin{array}{c}
               1 \\
               0\\
             \end{array}\right)\otimes...\otimes\left(\begin{array}{c}
               1 \\
               0\\
             \end{array}\right)\longrightarrow(P_{_{00...0}}=0,P_{_{00...01}}=0,P=1) \hspace{-2cm}\\
  \ket{\gamma}=\left(\begin{array}{c}
               1 \\
               0 \\
             \end{array}\right)\otimes\left(\begin{array}{c}
                1 \\
               0 \\
             \end{array}\right)\otimes...\otimes\left(\begin{array}{c}
               1 \\
               0 \\
             \end{array}\right)\longrightarrow(P_{_{00...0}}=1,P_{_{00...01}}=0,P=0) \\
            \ket{\gamma}=\left(\begin{array}{c}
               1 \\
               0 \\
             \end{array}\right)\otimes...\otimes\left(\begin{array}{c}
                1 \\
               0 \\
             \end{array}\right)\otimes\left(\begin{array}{c}
               0 \\
               1 \\
             \end{array}\right)\longrightarrow(P_{_{00...0}}=0,P_{_{00...01}}=1,P=0) \\
\end{array}\right.
$$
and consequently these yield the following hyperplanes surrounding
the feasible region (see Fig.2)
$$
P_{_{00...0}},P_{_{00...01}},P\geq0
$$
$$
1-P_{_{00...0}}- P_{_{00...01}}-P\geq0
$$
The positivity of the last constraint come from the positivity of
the  expectation value of the following optimal EW
$$
\mathcal{W}_{opt}=I-2|\psi_{_{00...0}}\rangle\langle
\psi_{_{00...0}}|-2 |\psi_{_{00...01}}\rangle\langle
\psi_{_{00...01}}|-\sigma
$$
in  pure product states $|\gamma\rangle$,  since it  can be
written as the partial transpose of a positive operator with
respect to the fist particle as
$$
\mathcal{W}_{opt}^{T_{1}}=2 (|\psi_{_{11...10}}\rangle\langle
\psi_{_{11...10}}| + |\psi_{_{11...11}}\rangle\langle
\psi_{_{11...11}}|).
$$
 Now the remaining task is to solve the following LP problem
$$
\mathrm{minimize} \quad\;\;a+(b-a)P_{_{00...0}}
 +(c-a)P_{
_{00...01}}+(d-a)P \hspace{1cm}
$$
\begin{equation}
 \mathrm{subject\ to} \quad\;\;\begin{array}{c}
  1-P_{_{00...0}}- P_{_{00...01}}-P\geq0 \hspace{3.5cm}\\
 P_{_{00...0}},P_{_{00...01}},P\geq0 \hspace{3cm}\\
\end{array}
\end{equation}
Thus,  above problem is reduced to LP and can be solved by simplex
method. Putting the apexes in Eq.(\ref{W1}) we deduce that $a,b,c,d$
should be  positive. Now, the operator $\mathcal{W}_{1}$ fulfills
the properties of EWs if at least one of its eigenvalues is
negative, namely $2b-a<0$ or $2c-a<0$.\\
The second example which sets in LP problem is
$$
\mathcal{W}_{2}=aI_{2^{n}}+2^{n-1} (b-a)\ket{\psi_{
_{00...0}}}\bra{\psi_{ _{00...0}}} +2^{n-1} (c-a)\ket{\psi_{
_{10...0}}}\bra{\psi_{ _{10...0}}}
$$
which  can not be solved by the prescription of this paper.  Its
feasible region can be determined by  Lagrangian multiplier
method, as it is discussed in   \cite{Jafar1}. Let $ P_{00...0}=2\
|\langle\psi_{_{00...0}}\ket{\gamma}|^{2}$ ,$P_{10...0}=2\
|\langle\psi_{_{10...0}}\ket{\gamma}|^{2}$. There the problem
reduces to
$$\hspace{-4cm}
\mathrm{minimize} \quad\;\;a+2^{n-2}(b-a)P_{_{00...0}}
 +2^{n-2}(c-a)P_{
_{10...0}}
$$
\begin{equation}
 \mathrm{subject\ to} \quad\;\;\left\{\begin{array}{c}
  \frac{1}{2^{n-2}}-P_{_{00...0}}+(1-\frac{1}{2^{n-2}})P_{_{10...0}}\geq0 \hspace{3.5cm}\\
 \frac{1}{2^{n-2}}-P_{_{10...0}}+(1-\frac{1}{2^{n-2}})P_{_{00...0}}\geq0 \hspace{3.5cm}\\
P_{_{00...0}},P_{_{10...0}}\geq0 \hspace{4cm}\\
\end{array}\right.
\end{equation}
These constraints can not be reach by partial transposition
approach, and  the feasible region is estimated by convex hall of
apexes (see Fig.3).

\section{Detecting some entangled states by $\mathcal{W}_{_{R}}^{(n)}$}
This section is devoted to  some entangled states which can be
detected by general $\mathcal{W}_{_{R}}^{(n)}$ and three-qubit
REW $\mathcal{W}_{_{R}}^{(3)}$.
 First  we consider some Bell states. All of the Bell states $\ket{\psi_{_{i\ 0...0}}}$ can be detected
by $\mathcal{W}_{_{R}}^{(n)}$, since we have
$$
Tr(\mathcal{W}_{_{R}}^{(n)}\ket{\psi_{_{00...0}}}\bra{\psi_{_{00...0}}})=\omega_{_{2}},
$$
$$
Tr(\mathcal{W}_{_{R}}^{(n)}\ket{\psi_{_{i\ 0...0}}}\bra{\psi_{_{i\
0...0}}})=\omega_{_{1}}\quad,\quad i\neq0,
$$
therefore for $\omega_{_{2}}<0$ one can detect
$|\psi_{_{00...0}}\rangle$
 and for  $\omega_{_{1}}<0$ the  others modulated
Bell states can be detected by $\mathcal{W}_{_{R}}^{(n)}$. On the
other hand, imposing some constraints on operator
\begin{equation}\label{den}
\rho_{_{i,0,...,0}}= \frac{1}{B\ Tr(\rho_{_{s}})+D
\;d_{1}}\left\{\;B( I - \sum_{j=0}^{d_{1}-1} \ket{\psi_{_{j\
0...0}}}\bra{\psi_{_{j\ 0...0}}})+ D \;d_{1}\ket{\psi_{_{i\
0...0}}}\bra{\psi_{_{i\ 00...0}}} \;\right\},
\end{equation}
one can get a density matrix which can be detected by
$\mathcal{W}_{_{R}}^{(n)}$, where $\rho_{_{s}}$ denotes the
separable state inside the parenthesis on the righthand side. The
positivity of $\rho_{_{i,0,...,0}}$ constrains  $B$ and $D$ to be
positive and in order to detect both $\rho_{_{0,0,...,0}}$ and
$\rho_{_{i,0,...,0}}, i\neq0$,  we must have
\begin{equation}\label{omeg2}
B\;
Tr(\mathcal{W}_{_{R}}^{(n)}\rho_{_{s}})+D\;d_{1}\;\omega_{2}<0,
\end{equation}
\begin{equation}\label{omeg1}
B\;
Tr(\mathcal{W}_{_{R}}^{(n)}\rho_{_{s}})+D\;d_{1}\;\omega_{1}<0,
\end{equation}
respectively. Because of the positivity of $D$, Eq.(\ref{omeg2})
is satisfied if $\omega_{2}<0$  and Eq.(\ref{omeg1}) is satisfied
if $\omega_{1}<0$. Now, we can proceed our discussion further to
detect PPT entangled states which is  useful for determining
non-decomposable region. Here we do not discuss the
decomposability and non-decomposability issues in detail, since it
needs the other opportunity and comes elsewhere.
$\rho_{_{i,0,...,0}}$ is PPT states with respect to any
subsystems of the particles, if $B-D$ be positive, so the
Eq.(\ref{omeg2}) yields
\begin{equation}
1\leq\frac{B}{D}<\frac{-d_{1}\;\omega_{2}}{Tr(\mathcal{W}_{_{R}}^{(n)}\rho_{_{s}})}\quad
\Longrightarrow\quad \omega_{2}<\varpi
\end{equation}
where
\begin{equation}
\varpi:=-\frac{1}{d_{1}}Tr(\mathcal{W}_{_{R}}^{(n)}\rho_{_{s}})
\end{equation}
The above requirement  makes $\rho_{_{0,0,...,0}}$  a class of PPT
entangled state which can be detected by
$\mathcal{W}_{_{R}}^{(n)}$ . But to detect  the other
$\rho_{_{i,0,...,0}}$ we must have $\omega_{1}<\varpi$ which is
impossible. This is in agreement  with the   discussion made in
section \ref{optimal}. Now, combining thus obtained results with
those of   Eq. (\ref{decom}) which implies that
$\mathcal{W}_{_{R}}^{(n)}$ is decomposable provided that
$\omega_{_{2}}$ is positive,  one  can rather determine the
decomposability and non-decomposability of
REWs (see Fig.4).\\
Furthermore  one can construct some entangled states which can be
detected by particular REWs. As an example consider entangled
density matrices
$$
\rho=\frac{1}{4B+2D}(B\ \sigma_{_{2}}+2\ D
|\psi_{_{000}}\rangle\langle\psi_{_{000}}|)
$$
$$
\rho'=\frac{1}{4B+2D}(B\ \sigma_{_{3}}+2\
D|\psi_{_{000}}\rangle\langle\psi_{_{000}}|)
$$
which can be detected by three-qubit REW (\ref{three-EW}), with
some constraints.
 The positivity of these states implies that $B,B+2D\geq0$ and
the positivity of $\rho^{T_{3}},\rho'^{T_{2}}$ is  achieved if
$B\pm D\geq0$. In order that,  $Tr(\mathcal{W}_{_{R}}^{(3)}\rho)$
to be negative we should have $
B(a_{_{2,3}}+a_{_{2}})+D\omega_{_{2}}<0$, where  we have two
possibilities: for $D>0 $ we  have $\omega_{_{2}}<0$ and for $D<0
$ we  have $\omega_{_{2}}>0$.
\section{  Positive maps}
As it is shown in \cite{Jamiolkowski}, there is a close connection
between the positive maps and the entanglement witnesses, i.e.,
the Jamio\l kowski isomorphism
\begin{equation}\label{Jamiol1}
d_{_{1}}d_{_{2}}...d_{_{n}}(I_{d_{1}...d_{n}}\otimes{\mathcal{E}}
)|\psi_{_{+}}\rangle\langle\psi_{_{+}}|=\mathcal{W}^{(1,2,...,2n)}_{d_{1},d'_{1},...,d_{n},d'_{n}}
\quad,\quad d_{i}\leq d'_{i}\quad ,\quad i=1,...,n
\end{equation}
\begin{equation}\label{Jamiol2}
\mathcal{E}(\rho)=Tr_{(1,3,...,2n-1)}\left[\ \mathcal{W}^{(1,2,...,2n)}_{d_{1},d'_{1},...,d_{n},d'_{n}}
(\rho^T\otimes I_{d'_{1}d'_{2}...d'_{n}})\ \right]
\end{equation}
where
\begin{equation}
|\psi_{_{+}}\rangle=\frac{1}{\sqrt{d_{1}d_{2}...d_{n}}}\sum_{i_{1}=0}^{d_{1}-1}\sum_{i_{2}=0}^{d_{2}-1}
 ... \sum_{i_{n}=0}^{d_{n}-1}
|i_{1}^{(1)}i_{1}^{(2)}\rangle |i_{2}^{(3)}i_{2}^{(4)}\rangle... |i_{n}^{(2n-1)}i_{n}^{(2n)}\rangle
\end{equation}
is the maximally entangled state in $
{\cal{H}}_{d_{1}}^{(1)}\otimes {\cal{H}}_{d'_{1}}^{(2)}
\otimes...\otimes
{\cal{H}}_{d_{n}}^{(2n-1)}\otimes{\cal{H}}_{d'_{n}}^{(2n)}
$ . Hence the Jamio\l kowski isomorphism is a
one-to-one mapping between the set of trace preserving quantum
operations
\begin{equation}\label{pmap}
\mathcal{E}:
\mathcal{H}_{d_{1}}^{(1)}\otimes\mathcal{H}_{d_{2}}^{(3)}\otimes...\otimes\mathcal{H}_{d_{n_{1}}}^{(2n-1)}
\longrightarrow
\mathcal{H}_{d'_{1}}^{(2)}\otimes\mathcal{H}_{d'_{2}}^{(4)}\otimes...\otimes\mathcal{H}_{d'_{n_{1}}}^{(2n)}
\end{equation}
and $d_{1}\times d^{\prime}\times... \times d_{n}\times
d^{\prime}_{n} $ EWs
 if $d_{i}\leq d'_{i}$ for $  i=1,2,..., n$.\\
Now, using the Jamio\l kowski isomorphism  (\ref{Jamiol2}), we try
to construct the positive maps connected with REWs. Evidently
tensor product of some EWs  is also an EW in higher dimension. To
be more precise  let
$\mathcal{W}^{(1,2,...,2n)}_{d_{1}d'_{1}...d_{n}d'_{n}}$ be an EW
acting on $ {\cal{H}}_{d_{1}}^{(1)}\otimes
{\cal{H}}_{d'_{1}}^{(2)}\otimes
...\otimes{\cal{H}}_{d_{n}}^{(2n-1)}\otimes{\cal{H}}_{d'_{n}}^{(2n)}$
then depending on  possible partition
\begin{equation}
n=n_{_{1}}+ n_{_{2}}+...+n_{_{m}}\ ,\quad n_{i}\geq 1
\end{equation}
one can construct  an  EW by tensor product of REWs as
\begin{equation}
\mathcal{W}^{(1,2,...,2n)}_{d_{1},d'_{1},...,d_{n},d'_{n}}=
\mathcal{W}^{(1,2,...,2n_{1})}_{d_{1},d'_{1},...,d_{n_{1}},d'_{n_{1}}}
\otimes\mathcal{W}^{(2n_{1}+1,...,2n_{1}+2n_{2})}_{d_{n_{1}+1},d'_{n_{1}+1},...,d_{n_{1}+n_{2}},d'_{n_{1}+n_{2}}}
\otimes...\otimes
\mathcal{W}^{(2n-2n_{m}+1,...,2n)}_{d_{n-n_{m}+1},d'_{n-n_{m}+1},...,d_{n},d'_{n}},
\end{equation}
then using   Jamio\l kowski isomorphism  (\ref{Jamiol2}), one can
obtain the corresponding   positive map. For instance,
considering the tensor product of $n$  REWs(corresponding to  the
partition $n=1+1+...+1$)
$$
\mathcal{W}^{(1,2,...,2n)}_{d_{1},d'_{1},...,d_{n},d'_{n}}=\bigotimes_{k=1}^n
\mathcal{W}_{d_{k},d'_{k}}^{(2k-1,2k)}
$$
 with
$$
\mathcal{W}_{d_{k},d'_{k}}^{(2k-1,2k)}=a_{_{1}}^{(2k-1,2k)}I+(a_{_{2}}^{(2k-1,2k)}-a_{_{1}}^{(2k-1,2k)})
|\psi_{_{00}}^{(2k-1,2k)}\rangle\langle\psi_{_{00}}^{(2k-1,2k)}|+
a_{_{1}}'^{(2k-1,2k)}\sigma_{_{1}}'^{(2k-1,2k)}
$$
acting on Hilbert space $ \mathcal{H}_{d_{k}}^{(2k-1)}\otimes
\mathcal{H}_{d'_{k}}^{(2k)}$ with $ \sigma_{_{1}}'^{(2k-1,2k)}$
given in Eq. (\ref{rho'}),   we  get the following positive map
$$
\mathcal{E}^{(n)}(\rho) = \sum_{S\subseteqq \{1,3,...,2n-1\}}
\Gamma_{_{S}} O_{_{S}}
$$
where
$$
\Gamma_{_{S}}=\prod_{j\in N\setminus S}(a_{_{2}}^{(2j-1,2j)}-
a_{_{1}}^{(2j-1,2j)}),
$$
$$
O_{_{S}}=\bigotimes_{j_{i}\in
S}\left(a_{_{1}}^{(2j_{i}-1,2j_{i})}I_{d_{j_{i}}}^{(j_{_{i}}+1)}
+a_{_{1}}'^{(2j_{i}-1,2j_{i})}\sum_{k=d_{j_{i}}}^{d_{j_{i}}'-1}|k\rangle\langle
k|\right)
 \otimes Tr_{j_{1}...j_{|S|}}
(\rho)
$$
and $ (j_{1}...j_{|S|})$ is the ordered of $S$. Choosing all
$a_{_{2}}=a_{_{1}}'=0$ and  $a_{_{1}}=1$ yields the generalized
reduction map introduced in \cite{Hall1}. As an  example for $n=2$
one can easily verify that
$$
\mathcal{E}^{(2)}(\rho)=Tr_{_{1,3}}[\
\mathcal{W}_{d_{1},d_{1},d_{2},d_{2}}^{(1,2,3,4)}(\rho^{T_{_{1,3}}}\otimes
I_{d_{1}d_{2}}^{(2,4)}) \ ]=
a_{_{1}}^{(1,2)}a_{_{1}}^{(3,4)}Tr(\rho) I_{d_{1}}^{(2)}\otimes
I_{d_{2}}^{(4)}
 +a_{_{1}}^{(1,2)}(a_{_{2}}^{(3,4)}-a_{_{1}}^{(3,4)})I_{d_{1}}^{(2)}\otimes Tr_{_{1}}(\rho)
$$
$$
+ a_{_{1}}^{(3,4)}(a_{_{2}}^{(1,2)}-a_{_{1}}^{(1,2)})
Tr_{_{3}}(\rho)\otimes
I_{d_{2}}^{(4)}+(a_{_{2}}^{(1,2)}-a_{_{1}}^{(1,2)})(a_{_{2}}^{(3,4)}-a_{_{1}}^{(3,4)})\rho
\hspace{2.3cm}
$$
$$
a_{_{1}}^{(1,2)}a_{_{1}}'^{(3,4)}Tr(\rho) I_{d_{1}}^{(2)}\otimes
\sum_{k=d_{2}}^{d_{2}'-1}|k\rangle\langle k|
+(a_{_{2}}^{(1,2)}-a_{_{1}}^{(1,2)})a_{_{1}}'^{(3,4)}Tr_{_{3}}(\rho)
\otimes \sum_{k=d_{2}}^{d_{2}'-1}|k\rangle\langle k|
$$
$$
+a_{_{1}}'^{(1,2)}(a_{_{2}}^{(3,4)}-a_{_{1}}^{(3,4)})\sum_{k=d_{1}}^{d_{1}'-1}|k\rangle\langle
k|\otimes Tr_{_{1}}(\rho)
+a_{_{1}}'^{(1,2)}a_{_{1}}^{(3,4)}Tr(\rho)\sum_{k=d_{1}}^{d_{1}'-1}|k\rangle\langle
k|\otimes I_{d_{2}}^{(4)}
$$
$$
+a_{_{1}}'^{(1,2)}a_{_{1}}'^{(3,4)}Tr(\rho)\sum_{k=d_{1}}^{d_{1}'-1}|k\rangle\langle
k| \otimes \sum_{k=d_{2}}^{d_{2}'-1}|k\rangle\langle k|
$$
Taking
$a_{_{2}}^{(1,2)}=a_{_{2}}^{(3,4)}=a_{_{1}}'^{(1,2)}=a_{_{1}}'^{(3,4)}=0$
and $a_{_{1}}^{(1,2)}=a_{_{1}}^{(3,4)}=1$ yields
\begin{equation}
 \mathcal{E}^{(2)}(\rho)=Tr(\rho)I_{d_{1}}^{(2)}\otimes  I_{d_{2}}^{(4)}-
I_{d_{1}}^{(2)}\otimes Tr_{_{1}}(\rho)-Tr_{_{3}}(\rho)\otimes
I_{d_{2}}^{(4)}+\rho
\end{equation}
and choosing $a_{_{2}}^{(1,2)}=a_{_{2}}^{(3,4)}=0$ and
$a_{_{1}}^{(1,2)}=a_{_{1}}^{(3,4)}=-a_{_{1}}'^{(1,2)}=-a_{_{1}}'^{(3,4)}=1$
we reach the new reduction positive map for different dimensions
$$
\mathcal{E}^{(2)}(\rho)= Tr(\rho) I_{d_{1}}^{(2)}\otimes
I_{d_{2}}^{(4)} -I_{d_{1}}^{(2)}\otimes Tr_{_{1}}(\rho)
-Tr_{_{3}}(\rho)\otimes I_{d_{2}}^{(4)}+\rho -Tr(\rho)
I_{d_{1}}^{(2)}\otimes \sum_{k=d_{2}}^{d_{2}'-1}|k\rangle\langle
k| +Tr_{_{3}}(\rho) \otimes
\sum_{k=d_{2}}^{d_{2}'-1}|k\rangle\langle k|
$$
$$
\hspace{2cm}+\sum_{k=d_{1}}^{d_{1}'-1}|k\rangle\langle k|\otimes
Tr_{_{1}}(\rho)
-Tr(\rho)\sum_{k=d_{1}}^{d_{1}'-1}|k\rangle\langle k|\otimes
I_{d_{2}}^{(4)}
+Tr(\rho)\sum_{k=d_{1}}^{d_{1}'-1}|k\rangle\langle k| \otimes
\sum_{k=d_{2}}^{d_{2}'-1}|k\rangle\langle k|
$$
These examples  show that  one can construct more positive maps by
making tensor product of EWs $ \mathcal{W}_{_{R}}^{(n)}$ in
arbitrary way provided that the dimensionality condition
$d_{i}\leq d'_{i}$ is satisfied.

\section{Conclusion}
The  generalized reduction type entanglement witnesses with
simplex feasible regions are introduced, where the EWs
corresponding to hyperplanes surrounding the feasible regions are
optimal. These REWs are of types that their manipulation is
reduced to LP problem and can be solved exactly by using the
simplex method. As it shown above, the REWs  are decomposable in
cases if their  second eigenvalue, namely  $\omega_{_{2}}$ becomes
positive while for negative values of  $\omega_{_{2}}$, the
decomposability or non-decomposability of REWs is still open for
debate. Also various other interesting issues remain unsolved,
such as keeping the REWs in realm of LP problems despite of
adding some other operators or entangled states to them.

 \vspace{1cm}
\setcounter{section}{0}
 \setcounter{equation}{0}
 \renewcommand{\theequation}{I-\arabic{equation}}

{\Large{Appendix I:}}\\
{\bf Proof of the inequalities:} $0\leq P_{S} , P'_{S}, P_{2,...,n}\leq1$.\\
 In this appendix we prove that all $P_{S}$ , $P'_{S}$ and
$P_{2,...,n}$ take the values between  0 and 1. The inequalities
$0\leq P_{S},P'_{S}\leq1$
 can be easily concluded from the following ones
$$
0\leq Tr( \sigma_{S} |\gamma\rangle\langle\gamma|)
\leq\prod_{k=1}^n \sum_{i=0}^{d_{k}-1}|\alpha_{i}^{(k)}|^2=1,
$$
$$
0\leq Tr( \sigma'_{S} |\gamma\rangle\langle\gamma|)
\leq\prod_{k=1}^n \sum_{i=0}^{d_{k}-1}|\alpha_{i}^{(k)}|^2=1.
$$

 For $P_{2,...,n}$,   the
Cauchy-Schwartz inequality implies that
$$
P_{2,...,n}:=d_{1}\
|\langle\psi_{_{00...0}}\ket{\gamma}|^{2}=\left|\sum_{i=0}^{d_{1}-1}\alpha_{_{i}}^{(1)}\alpha_{_{i}}^{(2)}...
\alpha_{_{i}}^{(n)}\right|^2 =
|\langle\alpha^{(1)}|\beta\rangle|^2\leq\||\alpha^{(1)}\rangle\|^2\||\beta\rangle\|^2,
$$
where
$$
|\beta\rangle=\left(\begin{array}{c}
                      \alpha_{0}^{(2)}\alpha_{0}^{(3)}...\alpha_{0}^{(n)} \\
                      \alpha_{1}^{(2)}\alpha_{1}^{(3)}...\alpha_{1}^{(n)} \\
                      \vdots \\
                      \alpha_{d_{1}-1}^{(2)}\alpha_{d_{1}-1}^{(3)}...\alpha_{d_{1}-1}^{(n)} \\
                    \end{array}
\right),
$$
finally using the following inequality
$$
\||\beta\rangle\|^2=\sum_{i=0}^{d_{1}-1}|\alpha_{i}^{(2)}\alpha_{i}^{(3)}...\alpha_{i}^{(n)}|^2
\leq\prod_{k=2}^n \sum_{i=0}^{d_{k}-1}|\alpha_{i}^{(k)}|^2=1,
$$
one can conclude that $0\leq P_{2,...,n}\leq1$.\\

{\Large{Appendix II:}}\\
{\bf Proof of the feasible region of (\ref{region}).}\\
In order to obtain the feasible region of (\ref{region}), we need
to evaluate the expectation value of  optimal  EWs

\begin{equation}
    ^{(S)}{\mathcal{W}_{_{opt}}^{(n)}}=a_{_{S}}\left(\sigma_{_{S}}
   +\sum_{_{ S\subsetneqq
  S'\neq N'}}(-1)^{|S|+|S'|}\sigma_{_{S'}}+d_{1}\;
(-1)^{|S|+|N'|}|\psi_{_{00...0}}\rangle\langle\psi_{_{00...0}}|-\sigma'_{_{S}}\right)
\quad S\subsetneqq N'
\end{equation}
in  pure product states $\ket{\gamma}\bra{\gamma}$  where
$^{(\emptyset)}{\mathcal{W}_{_{opt}}^{(n)}}=^{(1)}{\mathcal{W}_{_{opt}}^{(n)}}$.
Now by taking the partial transpose of
$^{(S)}\mathcal{W}_{_{opt}}^{(n)}$ with respect to $(N'\setminus
S)$ we have
\begin{equation}
 ^{(S)}{\mathcal{W}_{_{opt}}^{(n)}}^{T_{(N'\setminus S)}}=a_{_{S}}\sum_{i\neq
j}^{d_{1}-1}|\Psi_{ij}^{(S)}\rangle\langle\Psi_{ij}^{(S)}|
\end{equation}
where
\begin{equation}
|\Psi_{ij}^{(S)}\rangle:=|\alpha_{_{ij}}^{(1)}\rangle\otimes|\alpha_{_{ij}}^{(2)}\rangle\otimes...
\otimes|\alpha_{_{ij}}^{(n)}\rangle+
|\beta_{_{ij}}^{(1)}\rangle\otimes|\beta_{_{ij}}^{(2)}\rangle\otimes...\otimes|\beta_{_{ij}}^{(n)}\rangle,
\end{equation}
\begin{equation}
|\alpha_{_{ij}}^{(k)}\rangle
 = \left\{\begin{array}{c}
  |i\rangle\quad \quad\mathrm{if}\quad k\in 1\cup S \\
  |j\rangle   \quad\quad   \mathrm{if}\quad k\notin 1\cup S\\
\end{array}\right.\quad,\quad
|\beta_{_{ij}}^{(k)}\rangle
 = \left\{\begin{array}{c}
  |j\rangle\quad \quad\mathrm{if}\quad k\in 1\cup S \\
  |i\rangle   \quad\quad   \mathrm{if}\quad k\notin 1\cup S\\
\end{array}\right.
\end{equation}
Noting that all of these operators are positive definite and using
the relation
$$Tr(\
^{(S)}{\mathcal{W}_{_{opt}}^{(n)}}^{T_{(N'\setminus S)}}
\ket{\gamma}\bra{\gamma}\ )=Tr(\
^{(S)}{\mathcal{W}_{_{opt}}^{(n)}}
(\ket{\gamma}\bra{\gamma})^{T_{(N'\setminus S)}}\ )\geq 0
$$
yields all feasible regions which are simplexes.\\

{\Large{Appendix III:}}\\
{\bf  Proof of the optimality  of
$^{(S)}\mathcal{W}_{_{opt}}^{(n)}$. }\\
Here in this appendix we try to prove that  witness
$^{(S)}{\mathcal{W}_{_{opt}}^{(n)}}$ is optimal, to this aim  we
give the proof for the special case
$^{(2)}{\mathcal{W}_{_{opt}}^{(n)}}$, since the proof of general
case is rather similar to this particular one. According to the
Reference \cite{Lewen2}, the  EW
$^{(2)}{\mathcal{W}_{_{opt}}^{(n)}}$, is optimal if for all positive
operator P and $\varepsilon>0$, the following new Hermitian operator
\be\label{appp2} \mathcal{W}_{new}= (1+\varepsilon)
^{(2)}{\mathcal{W}_{_{opt}}^{(n)}}-\varepsilon P \ee is not anymore
an EW. Suppose that there is a positive operator $P$ and
$\epsilon\geq0$ such that
  $\mathcal{W}_{new}=\; ^{(2)}{\mathcal{W}_{_{opt}}^{(n)}}-\epsilon P $ is yet an EW.
 Let the positive operator $P$ be the pure
projection operator $|\psi_{i}\rangle\langle\psi_{i}|$, since an
arbitrary positive operator  can be written as  sum of pure
states with positive coefficients as
$P=\sum_{i}\lambda_{i}|\psi_{i}\rangle\langle\psi_{i}|$.

Now, one should note that the  expectation value of  the operator
$^{(2)}{\mathcal{W}_{_{opt}}^{(n)}}$ in pure product states $
|\gamma\rangle$ will vanish if they  satisfy the following
equation
 \begin{equation}\label{2}
 A_{i}B_{j}^*+A_{j}B_{i}^*=0
\end{equation}
with

$$
A_{i}=(\alpha_{_{1}})_{i}(\alpha_{_{2}})_{i}\quad\quad,\quad\quad
B_{j}=(\alpha_{_{3}})_{j}(\alpha_{_{4}})_{j}...(\alpha_{_{n}})_{j}.
$$
But, it is straightforward to see that, for $A_{i},B_{j}\in
\mathbb{R}$,  the pure state $|\psi\rangle\langle\psi|$ will be
similar to one of the
$|\Psi_{ij}^{(2)}\rangle\langle\Psi_{ij}^{(2)}|$ with $i\neq j$
with
\begin{equation}
|\Psi_{ij}^{(2)}\rangle:=|i\rangle\otimes|i\rangle\otimes|j\rangle\otimes...
\otimes|j\rangle+
|j\rangle\otimes|j\rangle\otimes|i\rangle\otimes...
\otimes|i\rangle \quad\quad i,j=0,...,d_{1}-1,
\end{equation}

concluding that an arbitrary $P$ has the form:
 $P=\sum_{i\neq j}^{d_{1}-1}a_{ij}|\Psi_{ij}^{(2)}\rangle\langle\Psi_{ij}^{(2)}|$ with $a_{ij}\geq0$.
 Finally, substituting Eq. (\ref{2}) in the
following expression
$$
Tr( P\ket{\gamma}\bra{\gamma}
)=\sum_{ij}a_{ij}|A_{i}B_{j}+A_{j}B_{i}|^2=
\sum_{ij}a_{ij}\left|\frac{A_{j}}{B_{j}}\right|^2|B_{j}^*B_{i}-B_{j}B_{i}^*|^2=0
$$
and choosing  $B_{i}$'s  such that $B_{j}^*B_{i}\neq B_{j}B_{i}^*$
 yields  $a_{ij}=0$ and  consequently one can conclude that
$P=0 $.

\newpage
{\bf Figure Captions}

 {\bf Figure-1:} 3-simplex displaying the
feasible region of three-qubit REW.

{\bf Figure-2:} 3-simplex displaying the feasible region of
multi-qubit  BDEW $\mathcal{W}_{1}$.

{\bf Figure-3:} Convex polygon displaying the boundaries of the
feasible region for multi-qubit BDEW $\mathcal{W}_{2}$.

{\bf Figure-4:}  Decomposable and non-decomposable regions of
REWs: for $\omega_{_{2}}\geq0$  the REWs are decomposable, for
$\omega_{_{2}}<\varpi$ the REWs are non-decomposable and in dashed
region, the  decomposability or non-decomposability of REWs is
still open for debate.


\begin{thebibliography}{99}

\bibitem{Preskill}
J. Preskill, {\it The Theory of Quantum Information and Quantum
Computation } \\ (California Inatitute of Technology, Pasadena, CA,
2000), http://www.theory.caltech.edu/poeole/preskill/ph229/.
\bibitem{Nielsen}
M. A. Nielsen and I. L. Chuang, {\it Quantum Computation and
Quantum Information} (Cambridge University Press, Cambridge,
2000).
\bibitem{Ekert}
{\it The Physics of Quantum Information}: Quantum Cryptography,
Quantum Teleportation and Quantum Computation, edited by D.
Bouwmeester, A. Ekert, and A. Zeilinger (Springer, New York,
2000).
\bibitem{Horod1}
M. Horodecki, P. Horodecki and R. Horodecki, Springer Tracts Mod.
Phys. \textbf{173}, 151 (2001).
\bibitem{Lewen1}
M. Lewenstein, D. Bru{\ss}, J.I. Cirac, B. Kraus, M. Kus, J.
Samsonowicz, A. Sanpera, and R. Tarrach, J. Mod. Opt. \textbf{47},
2841 (2000).
\bibitem{Terhal}
B. M. Terhal, Theor. Comput. Sci. \textbf{287}, 313 (2002) .
\bibitem{Bruss}
D. Bru\ss, J. Math. Phys. \textbf{43}, 4237 (2002).
\bibitem{Bell}
J. S. Bell, Physics (N.Y.) \textbf{1}, 195 (1964).
\bibitem{Peres}
A. Peres, Phys. Rev. Lett. \textbf{77}, 1413 (1996).
\bibitem{Horod2}
M. Horodecki, P. Horodecki, and R. Horodecki, Phys. Lett. A
\textbf{223}, 1 (1996).
\bibitem{Wootters1}
C. H. Bennett, D. P. DiVincenzo, J. A. Smolin, and W. K. Wootters,
Phys. Rev. A \textbf{54}, 3824 (1996).
\bibitem{Wootters2}
W. K. Wootters, Phys. Rev. Lett. \textbf{80}, 2245 (1998).
\bibitem{Lewen2}
M. Lewenstein, B. Kraus, J.I. Cirac, and P. Horodecki, Phys. Rev.
A \textbf{62}, 052310 (2000); ibid. \textbf{63}, 044304 (2001).
\bibitem{Jafar1}
M. A. Jafarizadeh, M. Rezaee, S. K. A. Seyed Yagoobi, Phys. Rev. A
\textbf{72}, 062106 (2005).
\bibitem{Jafar2}
M. A. Jafarizadeh, M. Rezaee, S. Ahadpour, Phys. Rev. A \textbf{74},
042335 (2006).
\bibitem{Stromer}
E. St\"{o}rmer, Acta Math. \textbf{110}, 233 (1963); S. L.
Woronowicz, Rep. Math. Phys. \textbf{10}, 165 (1976); M. D. Choi,
Proc. Sympos. Pure Math. \textbf{38}, 583 (1982).
\bibitem{Doherty1}
A. C. Doherty, P. A. Parrilo, and F. M. Spedalieri, Phys. Rev. Lett.
\textbf{88}, 187904 (2002).
\bibitem{Doherty2}
A. C. Doherty, P. A. Parrilo, and F. M. Spedalieri, Phys. Rev. A.
\textbf{69}, 022308 (2004).

\bibitem{Doherty3}
R. O. Vianna, A. C. Doherty, eprint quant-ph/0608095 (2006).
\bibitem{Boyd}
S. Boyd and L. Vandenberghe, \emph{Convex Optimization}, Cambridge
University Press, (2004).
\bibitem{Hall1}
W. Hall, Phys. Rev. A \textbf{72}, 022311 (2005)
\bibitem{Lewen3} A. Ac\'{i}n, D. Bru\ss, M.
Lewenstein,  and A. Sanpera, Phys. Rev. Lett. \textbf{87}, 040401
(2001).
\bibitem{Jamiolkowski}
A. Jamio\l kowski, Rep. Math. Phys. \textbf{3}, 275 (1972).
\end{thebibliography}
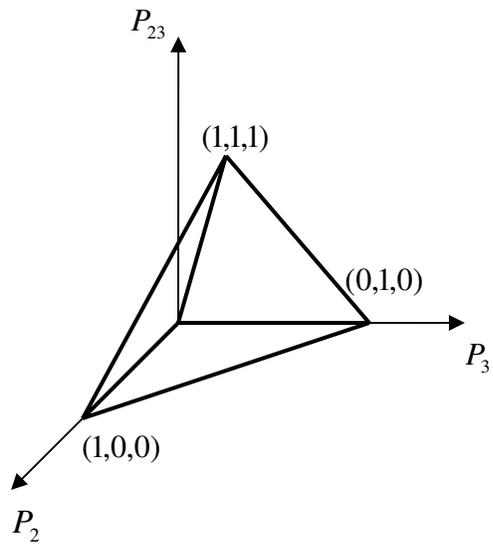
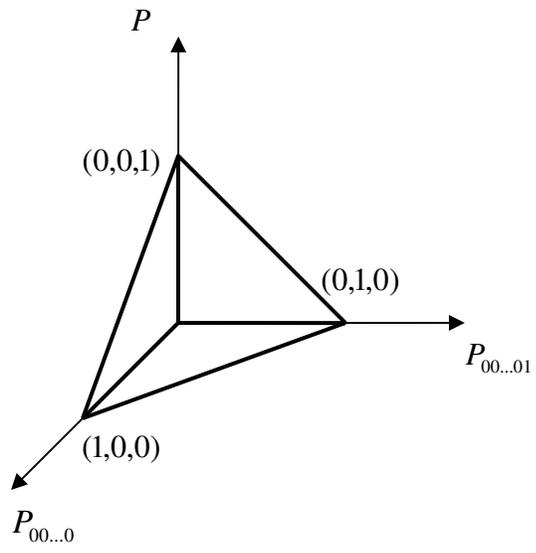
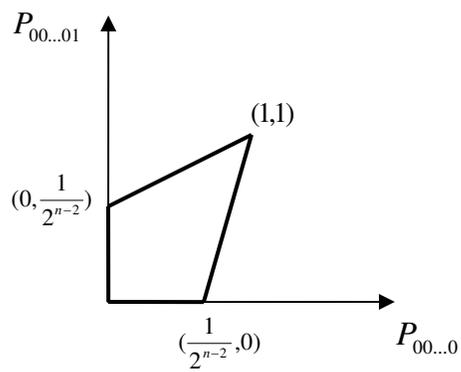
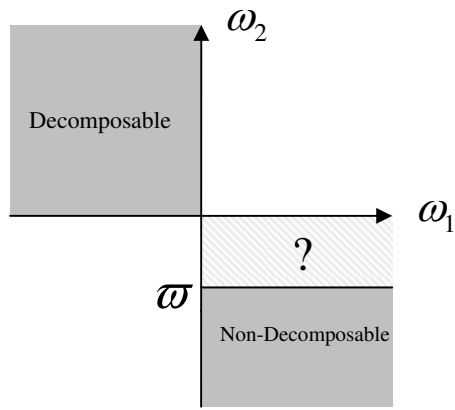
\end{document}